# A BAYESIAN APPROACH TO SOURCE SEPARATION

*Kevin H. Knuth*

Department of Neuroscience
Albert Einstein College of Medicine, Bronx NY 10461
kknuth@balrog.aecom.yu.edu

**ABSTRACT**

The problem of source separation is by its very nature an inductive inference problem. There is not enough information to deduce the solution, so one must use any available information to infer the most probable solution. We demonstrate that source separation problems are well-suited for the Bayesian approach which provides a natural and logically consistent method by which one can incorporate prior knowledge to estimate the most probable solution given that knowledge.

We derive the Bell-Sejnowski ICA algorithm from first principles, i.e. Bayes' Theorem and demonstrate how the Bayesian methodology makes explicit the underlying assumptions. We then further demonstrate the power of the Bayesian approach by deriving two separation algorithms that incorporate additional prior information. One algorithm separates signals that are known *a priori* to be decorrelated and the other utilizes information about the signal propagation through the medium from the sources to the detectors.

## 1. THE GENERAL METHODOLOGY

Typically one approaches a source separation problem with a set of mixed signals recorded by a set of detectors and some prior information regarding the physical situation in which these signals were emitted, propagated and recorded. Depending on the particular problem, this information may be limited or extremely detailed. Separation problems in which very little is known about the physical situation are typically called blind source separation problems. In these cases, prior information usually consists of the knowledge that the mixing is linear and that the amplitude densities of the source signals can be described by some class of probability densities.

Regardless of the specifics of the prior knowledge, there is rarely sufficient information from which one can deduce a unique solution. In these cases one must use a procedure of inductive reasoning to infer a solution. We will demonstrate that these problems can be effectively dealt with using the Bayesian methodology.

The general technique consists of forming a model that describes a particular source separation problem. The parameters describing a simple model may consist of the mixing matrix and the set of source signals, or may include more details such as the positions and orientations of the sources or their dynamic interactions. Once one has constructed a model consisting of a set of parameters that describes all of the relevant features of the source separation problem, one can calculate the probability that particular values of these parameters provide an accurate description the physical situation based on the acquired data and prior knowledge.

In this paper we apply the Bayesian methodology to the source separation problem and demonstrate how prior information can be incorporated to obtain a solution. We stress that the Bayesian interpretation of probability is quite different from the standard frequentist interpretation. Probability represents a degree of belief that a proposition is true and has nothing to do with treating our parameters as "random variables". We refer the interested reader to several works by E. T. Jaynes [1, 2, 3] and an excellent primer by Sivia [4].

## 2. BAYESIAN DERIVATION OF ICA

We focus on a source separation problem that can be described by a simple linear model consisting of a mixing matrix **A** and the source signal time series **s**(t), which is a vector composed of the time series describing the signals emitted by the individual sources. Bayes' Theorem is the natural starting point because it allows one to describe the probability of the model in terms of the likelihood of the data and the prior probability of the model and the data:

$$P(model \mid data, I) = \frac{P(data \mid model, I) P(model \mid I)}{P(data \mid I)}. \quad (1)$$

where $I$ represents any prior information. One can view Bayes' Theorem as describing how one's prior probability, $P(model \mid I)$, is modified by the acquisition of some new information.

To apply this to a source separation problem, we can consider the change in our knowledge about the system with the acquisition of new data consisting of mixtures of signals **x**(t) recorded by a set of detectors. In this case, Bayes' Theorem can be written as

$$P(\mathbf{A}, \mathbf{s}(t) \mid \mathbf{x}(t), I) = \frac{P(\mathbf{x}(t) \mid \mathbf{A}, \mathbf{s}(t), I) P(\mathbf{A}, \mathbf{s}(t) \mid I)}{P(\mathbf{x}(t) \mid I)}. \quad (2)$$

In most situations, one is not interested in calculating the probabilities. One usually wants to find the model that maximizes the probability in Equation (2). We can rewrite the equation as a proportionality and equate the inverse of the prior probability of the data $P(\mathbf{x}(t) \mid I)$ to the implicit proportionality constant

$$P(\mathbf{A}, \mathbf{s}(t) \mid \mathbf{x}(t), I) \propto P(\mathbf{x}(t) \mid \mathbf{A}, \mathbf{s}(t), I) P(\mathbf{A}, \mathbf{s}(t) \mid I). \quad (3)$$

The probability on the left-hand side of Equation (3) is referred to as the posterior probability. It represents the probability that a given model accurately describes the physical situation. The first term on the right-hand side is the likelihood of the data given the model. It describes the degree of accuracy with which we believe the model can predict the data. The final term on the right is the prior probability of the model, also called the prior. This prior describes the degree to which we believe the model to be correct based only on our prior information about the problem. It is through the assignment of the likelihood and priors that we express all of our knowledge about the particular source separation problem.

The prior probability can be factored into two terms since the mixing matrix, which describes the propagation of the signals from the sources to the detectors, is typically not affected by the amplitudes of the source signals. This results in

$$P(\mathbf{A}, \mathbf{s}(t) \mid \mathbf{x}(t), I) \propto P(\mathbf{x}(t) \mid \mathbf{A}, \mathbf{s}(t), I) P(\mathbf{A} \mid I) P(\mathbf{s}(t) \mid I). \quad (4)$$

The prior $P(\mathbf{A} \mid I)$ describes our prior knowledge regarding the form of the mixing matrix. This can include information about the propagation of the signals through the medium, the geometric arrangement of the detectors, or anything else that is known about the signal propagation and detection process. The prior $P(\mathbf{s}(t) \mid I)$ describes what is known about the source signals. Prior information about the amplitude density, frequency content, and dynamical behavior of the source waveforms are all represented by this prior.

If the linear mixing is relatively noise-free, it may be easier to estimate a separation matrix $\mathbf{W} = \mathbf{A}^{-1}$ that optimizes the posterior probability of the model and estimate the source signals by applying the separation matrix to the recorded data. In this case, we can treat the source signals as nuisance parameters and marginalize by integrating over all possible values of the source signals

$$P(\mathbf{A} \mid \mathbf{x}(t), I) \propto P(\mathbf{A} \mid I) \int d\mathbf{s}\ P(\mathbf{x}(t) \mid \mathbf{A}, \mathbf{s}(t), I) P(\mathbf{s}(t) \mid I). \quad (5)$$

The marginalization results in a posterior probability of the only model parameter of interest, the mixing matrix.

One finds that it is easier to estimate the values of the model parameters that maximize the posterior probability by looking at its logarithm. Taking the logarithm of Equation (5) one finds

$$\log P(\mathbf{A} \mid \mathbf{x}(t), I) =$$
$$\log \int d\mathbf{s}\ P(\mathbf{x}(t) \mid \mathbf{A}, \mathbf{s}(t), I) P(\mathbf{s}(t) \mid I) +$$
$$\log P(\mathbf{A} \mid I) + C. \quad (6)$$

Before assigning the specific probabilities that describe our prior knowledge, we preserve the generality of the derivation by first considering a stochastic gradient search technique to estimate the values of the separation matrix elements that maximize the logarithm of the posterior probability in Equation (6). The stochastic gradient method is an iterative process by which the values of the proposed separation matrix $\mathbf{W}$ are updated according to the derivative of Equation (6) with respect to $\mathbf{W}$ so that the value of the logarithm of the posterior probability increases until the global maximum is found. The update rule is written as

$$\mathbf{W}_{i+1} = \mathbf{W}_i + \Delta\mathbf{W} \quad (7)$$

where the matrix elements of $\Delta\mathbf{W}$ are found by

$$\Delta W_{ij} = \frac{\partial}{\partial W_{ij}} \log P(\mathbf{A} \mid \mathbf{x}(t), I)$$
$$= \frac{\partial}{\partial W_{ij}} \log \int d\mathbf{s}\ P(\mathbf{x}(t) \mid \mathbf{A}, \mathbf{s}(t), I) P(\mathbf{s}(t) \mid I) +$$
$$\frac{\partial}{\partial W_{ij}} \log P(\mathbf{A} \mid I). \quad (8)$$

It should be noted that $\Delta\mathbf{W}$, as defined by Equation (8), is not a matrix, but is the derivative of a scalar with respect to a matrix. The update rule can be made covariant by post-multiplying by $\mathbf{W}^T\mathbf{W}$ [5, 6].

We are now in a position to assign probabilities that express our prior knowledge about a particular source separation problem. In this example, we will make assumptions that lead to the Bell-Sejnowski ICA algorithm [7]. First, to express our belief that the mixing is linear, stationary and instantaneous, as described by $\mathbf{x}(t) = \mathbf{A}\mathbf{s}(t)$, and that the source signals are independent we assign a product of delta functions to the likelihood

$$P(\mathbf{x}(t) \mid \mathbf{A}, \mathbf{s}(t), I) = \prod_i \delta(x_i - A_{ik} s_k) \quad (9)$$

where we use the Einstein summation convention to denote matrix multiplication. If we have some reason to doubt that the mixing is described by the linear relation above, or we believe there may be noise present in the recordings, we can express this lack of confidence by assigning a prior that expresses a greater uncertainty, such as Gaussian density.

If we have some knowledge regarding the amplitude density of the source signals, we can assign an appropriate prior to describe this. As Bell and Sejnowski have shown, the amplitude densities of speech signals are well described by a hyper-Gaussian density. When assigning probabilities, one must be sure that the assignment accurately describes one's state of knowledge and does not presume more information than one possesses. For now, we denote the source amplitude probabilities by $p(s)$. The fact that the sources are believed to be independent is expressed by factoring the source signal prior

$$P(\mathbf{s}(t)|I) = \prod_l p_l(s_l). \quad (10)$$

Finally, the assumption of blind separation describes the fact that we know nothing about the form of the mixing matrix. This ignorance is represented by assigning a prior that is constant for all possible matrices $\mathbf{A}$ and zero outside of that range

$$P(\mathbf{A}|I) = \begin{matrix} C & \text{for all possible matrices} \\ 0 & \text{for all impossible matrices} \end{matrix} \quad (11)$$

As long as we constrain the search so that $\mathbf{A}$ is a possible mixing matrix, the prior is constant, and its derivative in Equation (8) is zero.

With the probabilities assigned, we outline the remainder of the derivation since it is identical to that performed by MacKay [6] and has been presented elsewhere [8, 9]. The multidimensional integral in Equation (8) describing the marginalization can be evaluated using the delta functions by introducing a change in variables, $w_i = x_i - A_{ik} s_k$,

$$\int d\mathbf{s} \ P(\mathbf{x}(t)|\mathbf{A},\mathbf{s}(t),I) P(\mathbf{s}(t)|I)$$
$$= \int d\mathbf{s} \ \prod_i \delta(x_i - A_{ik} s_k) \prod_l p_l(s_l)$$
$$= \frac{1}{\det \mathbf{A}} \prod_l p_l(A_{lk}^{-1} x_k). \quad (12)$$

Equation (8) which describes the update rule can be written as

$$\Delta W_{ij} = \frac{\partial}{\partial W_{ij}} \left[ -\log \det \mathbf{A} + \sum_l \log p_l(A_{lk}^{-1} x_k) \right]. \quad (13)$$

Recalling that $\mathbf{W} = \mathbf{A}^{-1}$, one can evaluate the derivative

$$\Delta W_{ij} = A_{ji} + x_j \left( \frac{p_i'(u_i)}{p_i(u_i)} \right)_j, \quad (14)$$

where $u_i = A_{ik}^{-1} s_k$. Equation (14) can be expressed in matrix form

$$\Delta \mathbf{W} = \mathbf{A}^T + \left( \frac{p_i'(u_i)}{p_i(u_i)} \right) \mathbf{x}^T, \quad (15)$$

which when made covariant by post-multiplying by $\mathbf{W}^T\mathbf{W}$ results in the Bell-Sejnowski ICA covariant update rule

$$\Delta \mathbf{W} = \mathbf{W} + \left( \frac{p_i'(u_i)}{p_i(u_i)} \right) \mathbf{u}^T \mathbf{W}. \quad (16)$$

It is important to note that the Bell-Sejnowski ICA algorithm derives from the assumptions of linear, stationary and instantaneous mixing, independence of the source signals and some basic information regarding the form of the source amplitude densities.

In previous work [6, 10], the model has been described at the outset to only consist of the mixing matrix. The source signals are then considered to be latent variables and appear in the calculation of the likelihood. Since the source signal prior does not explicitly appear, and one assigns a uniform density to the mixing matrix prior, the posterior probability consists only of the likelihood term. For this reason, ICA is considered to be a maximum likelihood approach.

For the sake of generality, we prefer to keep the priors explicit. While there is no discrepancy mathematically, we consider it methodologically important. It forces one to evaluate each prior separately and consider if there is anything known about the corresponding parameter, be it explicit or latent. For a given physical problem, one can often find additional information that can aid the search for a solution [9, 11, 12].

## 3. SEPARATING DECORRELATED SIGNALS

In this section we demonstrate the incorporation of additional information into a source separation problem. We consider a specific problem where it is known *a priori* that the signals to be separated are already decorrelated. In the case of artificially prewhitened data, this provides no additional information about the original problem. The explicit knowledge that the separation matrix must be orthogonal can still be used to constrain the form of the separation matrix. However, one should not expect this additional information to improve the accuracy of the results, since the assumption of statistical independence, which implicitly assumes decorrelation, has already been incorporated into ICA by factoring the source amplitude density prior. We have found that this additional information does constrain the solution and reduces the number of iterations necessary to attain convergence.

Prior information regarding the form of the mixing matrix is expressed by an appropriate assignment of the prior probability P($\mathbf{A}$ | I). We can express the fact that we believe the mixing matrix to be orthogonal in several ways. If this knowledge is very precise, we can assign a delta function to the prior. However, if we plan to use a stochastic gradient search we will do better using a prior having a continuous first derivative. As a candidate, we consider a Gaussian density

$$P(\mathbf{A}|I) = \frac{1}{\sqrt{2\pi}\sigma} Exp\left[ -\frac{\|\mathbf{A} - \mathbf{A}^{-T}\|^2}{2\sigma^2} \right] \quad (17)$$

where $\|\bullet\|$ denotes the Frobenius norm, which is the square root of the sum of squares of the matrix elements, **I** represents the identity matrix, *I* represents our prior information and $\sigma$ represents our uncertainty that **A** is orthogonal. Due to the symmetries of orthogonal matrices, one can find other arguments for the Gaussian prior, such as

$$P(\mathbf{A}|I) = \frac{1}{\sqrt{2\pi}\sigma} Exp\left[-\frac{\|\mathbf{I} - \mathbf{A}^T\mathbf{A}\|^2}{2\sigma^2}\right], \quad (18)$$

that describe prior probability densities that have maximal probability when **A** is orthogonal. We stress that that these priors are different and may give different results and "error bars" when used to estimate the maximum of the posterior probability. The assignment of a prior should be carefully based on the available information. The MaxEnt principle[†] can be used to show that the Gaussian prior in Equation (17) is appropriate when one has information regarding the expected squared deviation of **A** from $\mathbf{A}^{-T}$. While the prior in Equation (18) describes one's belief that the matrix is orthogonal, it has additional unknown biases. To illustrate the incorporation of this additional information, we will assign the former prior.

The mixing matrix prior in Equation (17) can be used in Equation (8) to represent our knowledge about the form of the mixing matrix. Taking the logarithm of the prior, we find the final term in Equation (8) to be

$$\frac{\partial}{\partial W_{ij}} \log P(\mathbf{A}|I) = \frac{\partial}{\partial W_{ij}}\left[-\frac{\|\mathbf{A} - \mathbf{A}^{-T}\|^2}{2\sigma^2}\right]. \quad (19)$$

The derivative can be performed by application of the chain rule and the observation that the Frobenius norm of a matrix $\|\mathbf{M}\|$ can be expressed as the trace of $\mathbf{M}^T\mathbf{M}$. The result is

$$\frac{\partial}{\partial W_{ij}} \log P(\mathbf{A}|I) =$$
$$\frac{1}{\sigma^2}\left[\mathbf{A}^T(\mathbf{A} - \mathbf{A}^{-T})\mathbf{A}^T - (\mathbf{A} - \mathbf{A}^{-T})^T\right]_{ij} \quad (20)$$

Adding this term to the previously obtained results in Equation (14) and post-multiplying by $\mathbf{W}^T\mathbf{W}$ to make the update rule covariant we get

$$\Delta\mathbf{W} = \mathbf{W} + \left(\frac{p'_i(u_i)}{p_i(u_i)}\right)\mathbf{u}^T\mathbf{W} +$$
$$\frac{1}{\sigma^2}\left[\mathbf{W}(\mathbf{W}^T - \mathbf{W}^{-1})\mathbf{W} - (\mathbf{W}^T - \mathbf{W}^{-1})^T\right]. \quad (21)$$

The last term is due to our explicit knowledge that **A** or equivalently, its inverse **W**, should be orthogonal. As one would expect, the term is zero when this is the case.

---
[†] See [4], p. 121

Preliminary simulations with variances ranging from 0.25 to 1.0 have shown that the algorithm converges with fewer iterations than ICA and that the accuracies of the solutions are comparable.

## 4. INVERSE SQUARE MIXING

In this section we demonstrate how more detailed information can be incorporated into a source separation problem. We consider an artificial problem where the signal amplitude decreases as the inverse square of the distance from the source. We also assume that the mean and variance of the source positions are known as well as a bound on the amplitude of the signals. Finally, as in ICA, we assume prior knowledge regarding the form of the source amplitude densities. This problem is described in more detail in a recent paper [9].

As in the previous section, our knowledge about the nature of the signal propagation from the sources to the detectors is equivalent to knowledge regarding the possible values of the mixing matrix elements. Specifically, in the case of an inverse square law we expect that

$$A_{ij} = \frac{a_j}{4\pi|r_{ij}|^2}, \quad (22)$$

where the value of the matrix element $A_{ij}$, is dependent on two parameters, the distance between the $i^{th}$ detector and the $j^{th}$ source and the amplitude $a_j$ of the source signal. The overall amplitude of the signal is explicitly included here since, like in ICA, the source amplitude density prior describes the statistics of the source signal, whereas the mixing matrix describes the overall amplitude of the signal and the transfer function. This is one source of the arbitrariness of the mixing matrix in ICA [8, 13].

We can use our knowledge about the prior probabilities of the source positions and the source amplitudes to determine a prior probability density for a given element of the mixing matrix. In this case we describe the probability of the amplitude of source *j* with a uniform density delimited by lower and upper cutoffs $b_{1j}$ and $b_{2j}$, respectively. This uniform density expresses our ignorance about the probability of observing amplitudes within that range. Given the mean value for the possible location of source *j* and a corresponding expected squared deviation from the mean, the MaxEnt principle insists that we assign a Gaussian density to the prior describing the source position. From this Gaussian density we can derive a MaxEnt probability for the source being situated a distance $|r_{ij}|$ from detector *i*. For the purposes of simplification, we utilize a Gamma prior to describe this prior probability:

$$P(|r_{ij}||I) = \frac{\beta^{-\alpha} Exp\left[-\frac{|r_{ij}|}{\beta}\right]|r_{ij}|^{-1+\alpha}}{\Gamma(\alpha)}, \quad (23)$$

where the mean distance between the detector and the source is given by $|\bar{d}_i - \bar{s}_j| = \alpha\beta$ and the expected squared deviation of the distance from the mean is given by $\sigma_j^2 = \alpha\beta^2$. From the uniform source amplitude prior and the Gamma distance prior, we can write the prior probability of an element of the mixing matrix as

$$P(A_{ij} | I) = \frac{4\pi}{\Gamma(\alpha_{ij})\beta_{ij}^2 (b_{2j} - b_{1j})} \times \left[ \gamma\left(2 + \alpha_{ij}, \frac{1}{\beta_{ij}}\sqrt{\frac{b_{1j}}{4\pi A_{ij}}}\right) - \gamma\left(2 + \alpha_{ij}, \frac{1}{\beta_{ij}}\sqrt{\frac{b_{2j}}{4\pi A_{ij}}}\right) \right] \quad (24)$$

where $\Gamma(\bullet)$ is the gamma function, $\gamma(\bullet, \bullet)$ is the incomplete gamma function [14], $\alpha_{ij}$ and $\beta_{ij}$ are found as described above and refer to the distance between detector $i$ and source $j$, and $b_{1j}$ and $b_{2j}$ are the minimum and maximum possible values, respectively, of the source amplitude. To obtain the necessary prior describing our expectations for the entire matrix, we make the simplifying assumption that information about one matrix element provides no information about the others. This is obviously not true when the geometric arrangement of the detectors is known. The mixing matrix prior is given by

$$P(\mathbf{A} | I) \approx \prod_{i,j} P(A_{ij} | I) \quad (25)$$

and Equation (24) above.

Now introducing this prior into Equation (8), we obtain an additional term describing the effect of our prior knowledge about the propagation of the signals on the update rule. Taking the derivative of the logarithm of the posterior probability and post-multiplying by $\mathbf{W}^T\mathbf{W}$ we again obtain an update rule similar to that used by ICA

$$\Delta\mathbf{W} = \mathbf{W} + \left(\frac{p_i'(u_i)}{p_i(u_i)}\right)\mathbf{u}^T\mathbf{W} - \mathbf{W}^{-T}\mathbf{M}\mathbf{W} \quad (26)$$

where

$$M_{ij} = \frac{B^D e^{-B} - C^D e^{-C}}{2 A_{ij} [\gamma(D,B) - \gamma(D,C)]} \quad (27)$$

with

$$B = \frac{1}{\beta_{ij}}\sqrt{\frac{b_{1j}}{4\pi A_{ij}}}, \quad C = \frac{1}{\beta_{ij}}\sqrt{\frac{b_{2j}}{4\pi A_{ij}}}, \quad D = 2 + \alpha_{ij}. \quad (28)$$

As a demonstration of the effectiveness of the additional prior information, we applied the algorithm to five artificially mixed sounds [9]. We have previously shown that two of the sounds could not be separated by ICA using a hyper-Gaussian source amplitude density prior, resulting in a diagonal solution[†]. However, the additional accurate information allowed this modified algorithm to separate the two source signals even though the source amplitude density prior presumed incorrect information about the source signals. Overall the separated signals contained slightly more noise than ICA, but the diagonal solution was averted. We stress that additional prior information about the mixing is not a substitute for an inaccurate representation of prior information regarding source amplitude densities.

---
[†] See [7] for a discussion on diagonal solutions.

## 5. CONCLUSIONS

We have demonstrated the application of the Bayesian methodology to three source separation problems. By re-deriving the Bell-Sejnowski ICA algorithm using the Bayesian approach we make explicit the assumptions and prior information incorporated into the problem. This provides some insight into the importance and meaning of the various terms in the ICA update rule. The methodology provides a natural and logically consistent means by which additional information can be incorporated. This was first demonstrated in the particular situation of the separation of decorrelated sources where the prior information consisted of a symmetry property of the mixing matrix. The final derivation demonstrated the inclusion of detailed information about a specific physical situation.

The Bayesian methodology has several advantages. First, all of the assumptions that go into finding a solution are made explicit. This is essential, since many *ad hoc* algorithms have implicit assumptions that may restrict the usefulness of the technique in unpredictable ways. It also forces the researcher designing the algorithm to consider the validity of each assumption.

Second, all of the prior knowledge about a specific problem is expressed in terms of prior probabilities that must be evaluated. This provides one with the means to incorporate any additional relevant information into a problem. Often something as simple as symmetry can provide a valuable constraint. It is sometimes the case that inclusion of what appears to be minimal information, such as the form of the source amplitude density, can be a powerful asset. In these cases, a simple algorithm can exhibit surprising success.

The Bayesian methodology is useful in cases where a given algorithm fails to produce accurate results. These events are interesting to a Bayesian since the implication is that there are flawed assumptions or incorrect prior information. One should consider the failure of an algorithm as a warning that our understanding of the situation is inadequate or incorrect. There is a close analogy between this viewpoint and the scientific method. The failure of a theory to predict an observation is an exciting event in that it provides the means to improve one's knowledge. In this sense, the failure of a separation algorithm in a particular situation provides the opportunity to improve the algorithm. Without the fact that the assumptions and priors are explicit, this improvement can be difficult indeed.

The explicit nature of the assumptions and priors also facilitates the generalization of algorithms to new domains of application. This is extremely important since it is often difficult to generalize *ad hoc* algorithms to other applications. Methodology is often viewed as trivial, especially when different methods lead to the same results. Jaynes expressed the importance of rationale aptly when he wrote, "Indeed, if we were to stay forever on the current problems, different rationales would be just different personal tastes without real consequences. But different rationales generalize differently."[‡] It is our view that this emphasis on methodology is important especially considering the success of ICA and the desire to utilize source separation techniques in other areas of application.

We hope to have demonstrated the relative ease with which the Bayesian methodology can be used to generalize a source separation algorithm to other applications by incorporating additional information such as a straightforward symmetry present in a problem or detailed knowledge of a specific physical situation.

## 6. REFERENCES

[1] E. T. Jaynes, "Bayesian methods: general background. An introductory tutorial", in *Maximum Entropy and Bayesian Methods in Applied Statistics*, J. H. Justice (ed.), Cambridge University Press, pp. 1-25, 1985.

[2] E. T. Jaynes, *Papers on Probability, Statistics and Statistical Physics*, R. D. Rosenkrantz (ed.), Dordrecht: D. Reidel Publishing Co., 1983.

[3] E. T. Jaynes, *Probability Theory - The Logic of Science*, unpublished, available at:
ftp://bayes.wustl.edu/pub/Jaynes/book.probability.theory/

[4] D. S. Sivia, *Data Analysis. A Bayesian Tutorial*, Oxford: Clarendon Press, 1996.

[5] S. Amari, "Natural gradient works efficiently in learning", *Neural Comp*, vol. 10, pp. 251-276, 1998.

[6] D. J. C MacKay, "Maximum likelihood and covariant algorithms for independent component analysis", Draft Paper, 1996, available at:
http://wol.ra.phy.cam.ac.uk/mackay/

[7] A. J. Bell and T. J. Sejnowski, "An information-maximization approach to blind source separation and deconvolution", *Neural Comp*, vol. 7, pp. 1129-1159, 1995.

[8] K. H. Knuth, "Difficulties applying recent blind source separation techniques to EEG and MEG", in *Maximum Entropy and Bayesian Methods, Boise 1997*, G. J. Erickson (ed.), Dordrecht: Kluwer Academic Publishers, To be published 1998.

[9] K. H. Knuth, "Bayesian source separation and localization", in *Proceedings of SPIE: Bayesian Inference for Inverse Problems*, vol. 3459, A. Mohammad-Djafari (ed.), To be published 1998.

[10] J.-F. Cardoso, "Infomax and maximum likelihood for blind source separation", *IEEE Signal Processing Letters*, vol. 4, no. 4, pp. 112-114, 1997.

[11] B. A. Pearlmutter and L. C. Parra, "A context-sensitive generalization of ICA", 1996 International Conference on Neural Information Processing, Hong Kong, 1996.

[12] K. H. Knuth and H. G. Vaughan, Jr., "Convergent Bayesian formulations of blind source separation and electromagnetic source estimation", in *Maximum Entropy and Bayesian Methods, Munich 1998*, Dordrecht: Kluwer Academic Publishers, To be published 1998.

[13] H. H. Yang and S. Amari, "Adaptive on-line learning algorithms for blind separation - maximum entropy and minimum mutual information", *Neural Comp*, vol. 9, no. 7, pp. 1457-1482, 1997.

[14] M. Abramowitz and I. A. Stegun, *Handbook of Mathematical Functions*, NY: Dover Publications, p. 260, 1972.

## ACKNOWLEDGEMENTS

This work was supported by NIH 5P50 DC00223, NIH 5P30 HD01799, and NIH NIDCD 5 T32 DC00039-05.

---

[‡] [1] p. 17.